\documentclass[a4paper,12pt]{article}
\usepackage[cp1251]{inputenc}
\usepackage[english]{babel}
\usepackage[T2A]{fontenc}
\usepackage{amsmath}
\usepackage{amssymb}
\usepackage{graphicx}

\hoffset=+5mm \voffset=+3mm \Roman{table}

\newcounter{tapp}

\begin{document}

\begin{table}[t]
\begin{tabular}{|p{17cm}|}
\hline
\it{Astronomy Letters, 2012, vol. 38, No. 5, pp. 290-304} \\
Translated from Pis'ma v Astronomicheskii Zhurnal, 2012, vol. 38,
No. 5, pp. 333-347 \\
\hline
\end{tabular}
\end{table}

\begin{center}
{\Large \bf{Study of the structure and kinematics of
the NGC~7465/64/63 triplet galaxies}}

\vspace{1cm}
     {\bf O.A. Merkulova\footnote{E-mail: olga\_merkulova@list.ru}$^{,\hspace{0.5mm}a}$,
     G.M. Karataeva$^a$, V.A. Yakovleva$^a$, A.N. Burenkov$^b$}

     { $^a$Astronomical Institute of Saint-Petersburg State University \\
       $^b$Special Astrophysical Observatory, Russian Academy of Sciences}

\abstract{
This paper is devoted to the analysis of new observational data
for the group of galaxies NGC~7465/64/63, which were
obtained at the 6-m telescope of the Special Astrophysical Observatory
of the Russian Academy of Sciences (SAO RAS) with
the multimode instrument SCORPIO and the Multi Pupil Fiber Spectrograph.
For one of group members (NGC~7465) the presence of a polar ring was suspected.
Large-scale brightness distributions, velocity
and velocity dispersion fields of the ionized gas
for all three galaxies as well as line-of-sight velocity curves
on the basis of emission and absorption lines and
a stellar velocity field in the central region for NGC~7465 were constructed.
As a result of the analysis of the obtained information,
we revealed an inner stellar disk (r $\approx$ 0.5~kpc)
and a warped gaseous disk in addition
to the main stellar disk, in NGC~7465.
On the basis of the joint study of
photometric and spectral data it was ascertained that
NGC~7464 is the irregular galaxy of the IrrI type,
whose structural and kinematic peculiarities resulted most likely from
the gravitational interaction with NGC~7465.
The velocity field of the ionized gas of NGC~7463
turned out typical for spiral galaxies with a bar, and
the bending of outer parts of its disk could arise owing to
the close encounter with one of galaxies of the environment.

 \it{Key words: galaxies, groups of galaxies,
 interacting galaxies -- kinematics, structure}
}
\end{center}

\section{Introduction}

The NGC~7465/64/63 triplet belongs to the NGC~7448 group that
consists of 9 galaxies with line-of-sight velocities
from 1800 to 2350~km/s and with the velocity variance of
${\sim}150$~km/s (Li and Seaquist, 1994). A summary of main
properties of the NGC~7465/64/63 galaxies (which were taken
from other papers and obtained during our study) is given
in Table~\ref{t:data7465}. Both individual galaxies and the whole group
have been studied intensively. The extensive investigation
\begin{table}
\begin{center}
\caption{Main characteristics of the NGC~7465/64/63 galaxies}
\label{t:data7465}
\vskip0.1cm
\begin{tabular}{c|c|c|c|c|c|c}
\hline \hline
Characteristics & NGC~7465 & Refer. & NGC~7464 & Refer. & NGC~7463 & Refer.\\
\hline
Morphological type & (R$'$)SB0$^0$ & RC3 & E1 pec       & RC3   & SABb: pec & RC3 \\
              &  Sab?         &  1  & IrrI         & 1     &           &      \\
$V_{hel}$, km/s& 1963 $\pm$ 3 & 1 & 1765 $\pm$ 2 & 1 & 2366  $\pm$ 3 & 1 \\
B$_{t,0}$, mag& 13.05 & 1 & 14.23 & 1 & 12.35 & RC3\\
(B-V)$_0$, mag& 0.85 & 1 & 0.40 & 1 & 0.28 & RC3\\
(V-R)$_0$, mag& 0.47 & 1 & 0.32 & 1 & & \\
$M_B$, mag& --19.30 & 1 & --18.12 & 1 & --19.90 & LEDA\\
F$_{H\alpha}^*$, erg/s/cm$^2$ & 1.1 $\times$ 10$^{-12}$ & 1 &  &   &   &  \\
SFR$_{H\alpha}$, $M_{\odot}$/year & 0.9 & 1 &  &   &   &  \\
$V_{max}$, km/s& 105 & 1 & 40 & 1 & 115 & 1 \\
$V_{hel}$ (HI data), km/s& 1962 $\pm$ 6 & LEDA & 1870 $\pm$ 9 & LEDA & 2445 $\pm$ 8 & LEDA \\
F$_{\lambda21cm}$, mag& 14.53 $\pm$ 0.16 & LEDA & 14.66 $\pm$ 0.15 & LEDA & 15.33 $\pm$ 0.16 & LEDA\\
log$\frac{M_{HI}}{M_{\odot}}$ & 9.70 $\pm$ <0.01 & 2 & 9.3 & 3 & 9.2 & 3 \\
$W_{50}$, km/s& 81 & 3 & 292 & 4 & 218 & 4 \\
\hline \hline
\end{tabular}
\end{center}
\vskip0.3cm

{\footnotesize
1 -- this paper,
2 -- Fernandez et al. (2010),
3 -- van Driel et al. (1992),
4 -- Springob et al. (2005)

*F$_{H\alpha}$ is the total flux within the
9.3 $\times$ 10$^{-19}$~erg/s/cm$^2$/arcsec$^2$ isophote,
SFR$_{H\alpha}$ is the corresponding star formation rate.}

\end{table}
of six members of the NGC~7448 group in the optical,
far infrared and radio ranges (van Driel et al., 1992)
showed that most of these galaxies are peculiar.
The undisturbed HI distribution and kinematics were
revealed in NGC~7448, NGC~7463, UGC~12313, and UGC~12321.
In optical images of NGC~7448 and NGC~7463, distortions
of outer spiral arms are seen well; NGC~7464 и NGC~7465
have been identified as merger candidates because of
their large ultraviolet (UV) excesses, the intense
star formation regions, etc. In the optical range,
all objects show emission lines; most of the galaxies
have HII-type spectra indicating the active star formation;
some galaxies rather resemble LINERs. The nuclear activity
is observed only in the NGC~7465 galaxy. NGC~7465 together
with NGC~7464 and NGC~7463 form a close subgroup.

NGC~7464 and NGC~7465 (Mrk~313) are early-type galaxies
with the projected distance between them of
$1\overset{\prime}{.}9$, and NGC~7463 is classified
as a peculiar barred spiral galaxy seen almost edge-on.
It is ${\sim}0\overset{\prime}{.}7$ distant from NGC~7464
and $2\overset{\prime}{.}5$ distant from NGC~7465.
These three galaxies have been classified as
UV-excess galaxies (Takase, 1980). NGC~7465 was
included in the catalog of Whitmore et al. (1990)
in the D category (D-42) --- the category of systems
possibly related to polar-ring galaxies (PRGs) ---
because of a faint outer ring at a position angle of
${\approx}45^{\circ}$ that is seen in the optical range.

High resolution 21-cm observations of this compact triplet
and another two faint galaxies (UGC~12313 and UGC~12321)
from the NGC~7448 group (sometimes these 5 galaxies are
united into the small NGC~7465 group) were obtained
by Li and Seaquist (1994). It was shown in their paper
that tidal disturbances in the HI morphology and kinematics
are observed in four of five members of the NGC~7465 group.
It was ascertained that an HI ``bridge'' connects, in projection,
UGC~12313 with NGC~7463 or with the triplet as a whole;
and sides of both galaxies connected by the ``bridge''
show signs of the recent interaction. Near the center of NGC~7463,
the change in the position angle of the HI disk is observed,
and in the same region the velocity field reverses its sign.
These peculiarities are explained by the existence of
the ``high-velocity'' gas, most of which is above
the eastern side of the galaxy's optical disk.
The HI emission of NGC~7463 is separated from that of
NGC~7465/64 in the velocity space. To the south and southeast
of NGC~7465 at a distance of ${\sim}80''$ (11.4 kpc),
an arc-like structure is observed. Together with some weak
emission to the north of NGC~7465, they form a ring.
Taking into account the orientation of NGC~7465 and
the HI ring, authors of the mentioned work suggest
that it represents a polar ring around NGC~7465 and
could be material pulled out from NGC~7464 during
a close encounter with NGC~7465. The southern part of
the ring that contains most of the HI flux coincides
with the string of condensations seen in optical images
at a level of ${\sim}24$~mag arcsec$^{-2}$ in the $V$ band
(Casini and Heidmann, 1978; van Driel et al., 1992).
However, the optical string apparently extends further
to the southwest rather than bends toward NGC~7464,
as the HI ring does. Therefore, it is not clear if
the optical string and the HI ring are related.

Since the panoramic (2D) spectroscopy makes it possible
to obtain a more complete kinematical picture of an object,
especially in the case of the presence of several components,
the peculiar galaxies NGC~7465/64/63 were included in
the program of our investigation. This paper is the
continuation of the study of candidates for PRG by the
2D-spectroscopy methods that was started in the paper
of Shalyapina et al. (2007). In addition to spectral
observations, we tried to obtain deep images in the
optical range for a more detailed study of the morphology
and structure of these objects. In the next section
the brief information about instruments used for observations
and about methods of the data processing is given.
Then the results of our study of the structure and
kinematics of each galaxy are presented. In conclusion
the discussion of all available data is carried out.

As distances to the NGC~7465/64/63 galaxies, we take
the distance to the NGC~7448 group of 29.5~Mpc
($H_0 = 75$~km/s/Mpc) (van Driel et al., 1992);
then a scale is: $1'' = 0.14$~kpc.

\section{Observations and processing}

Observations of the NGC~7465/64/63 galaxies were performed
at the 6-m telescope of SAO RAS. The detector was
the EEV 42-40 CCD-array of $2048 \times 2048$ pixels
(each pixel size is $13.5 \times 13.5~\mu m$).

Photometric observations of the galaxies in
the Johnson $B$ and $V$ bands and in the Cousins $R$ band
were performed using the SCORPIO focal reducer
(Afanasiev and Moiseev, 2005) at the night of 16 to 17
August, 2004. For the calibration, standard stars
from a list of Landolt (1983) were observed during
the night. The information about the photometric
observations is given in Table~\ref{t:zhurnal_phot}.
\begin{table}[hp]
\begin{center}
\caption{Photometric observations}
\label{t:zhurnal_phot} \vskip0.3cm
\begin{tabular}{c|c|c|c}
\hline \hline
Object & Band & Exposure time  & z, deg \\
       & & (frames~$\times$~s) & \\
\hline
NGC~7465/64/63 & $B$ & 600 + 2$\times$300 + 2$\times$30 & 40--44 \\
        & $V$ & 9$\times$60 & 37--40 \\
         & $R_c$ & 4$\times$30 + 2$\times$20 + 3$\times$120 & 35--37 \\
\hline \hline
\end{tabular}
\end{center}
\end{table}
The observations
were processed using the ESO-MIDAS software package.
Transparency coefficient values average for SAO RAS
(Neizvestnyi, 1983) were used in the reduction for
the atmosphere. An accuracy of total magnitude
estimates of galaxies is $\pm0^m.1$.

Spectral observations of the galaxies were also
carried out at the prime focus of the 6-m telescope,
using the SCORPIO focal reducer in modes of the
Interferometer Fabry-Perot (IFP) or of the long-slit
spectroscopy (LS) and using the Multi Pupil Field
Spectrograph (MPFS) (see the SAO RAS web
site\footnote{http://www.sao.ru/hq/lsfvo/devices/mpfs/mpfs\_main.html},
Afanasiev et al., 2001). A log of observations of
the galaxies is given in Table~\ref{t:zhurnal_spectr}.
\begin{table}[h!]
\begin{center}
\caption{Spectral observations}
\label{t:zhurnal_spectr}
\vskip0.3cm
\begin{tabular}{c|c|c|c|c|c|c}
\hline \hline
Object & Instrument, & Exposure & Field & Seeing, & Spectral    & Р.А. \\
       & data        & time, s  &       & arcsec  & region, \AA & ~ \\
\hline
NGC~7465 & LS 16.08.2006 & 4$\times$1200 & 1$''\times$ 6$'$ & 1.8 & 4800--5570 & 160$^\circ$ \\
NGC~7465/64/63 & IFP 16.08.2006 & 32$\times$180 & 6$'\times$ 6$'$ & 1.8 & H$\alpha$ & \\
NGC~7465 & LS 26.07.2008 & 2$\times$900 & 1$''\times$ 6$'$ & 1.6 & 5700--7400 & 161$^\circ$ \\
NGC~7465 & LS 27.07.2008 & 2$\times$1200 & 1$''\times$ 6$'$ & 2.9 & 5700--7400 & 45$^\circ$ \\
NGC~7465 & MPFS 05.08.2008 & 3$\times$900 & 16$''\times$ 16$''$ & 2.0 & 4196--5712 & Central \\
         &                 &              &                     &     &            & region\\
NGC~7465 & MPFS 08.08.2008 & 4$\times$900 & 16$''\times$ 16$''$ & 3.0 & 5630--7166 & Central\\
         &                 &              &                     &     &            & region\\
\hline \hline
\end{tabular}
\end{center}
\end{table}

Parameters of the focal reducer during the interferometric
observations are given in Moiseev (2002). The preliminary
monochromatization was carried out using the narrow-band
filter IFP 661 with a central wavelength
$\lambda_{\textrm{c}}$ of 6604\AA ~and a full width at
half maximum $\textrm{FWHM} = 21$~\AA. The spacing
between adjacent orders of interference was 28~\AA~ (1270~km/s).
The spectral resolution of the IFP was 2.5~\AA~ (${\approx} 110$~km/s).
The readout of the detector (EEV 42-40 CCD-array) was
carried out in the mode of $4 \times 4$ pixel hardware averaging,
so $512\times512$ pixel images (the pixel size was $0.714''$)
were obtained in each spectral channel.

The processing of interferometric observations was carried out
using the software developed at SAO RAS (Moiseev, 2002).
After primary procedures (the subtraction of night-sky lines
and the reduction to the wavelength scale), the observational
data represented ``data cubes'', where each point in
a $512 \times 512$-element field contained a 32-channel
spectrum. The optimal data filtering --- a Gaussian
smoothing over spectral coordinate with FWHM equal to
1.5 channels and the two-dimensional Gaussian smoothing
over spatial coordinates with FWHM = 2 pixels ---
was carried out. The Gaussian fitting of the H$\alpha$
and [NII]~$\lambda$6584~\AA emission line profiles was
used to construct velocity fields and monochromatic images.
Measurement errors of line-of-sight velocities for lines
with symmetric profiles were about 10~km/s. We also
constructed images in the 6592.0--6619.35~\AA
(near H$\alpha$) continuum.

For the analysis of a velocity field we used
the ``tilted-ring'' method (Begeman, 1989; Moiseev,
Mustsevoi, 2000); it makes it possible to determine
a dynamical center position and a position angle of
dynamical axis, to refine an inclination of a galaxy
to the plane of the sky, and to construct a rotation curve.
The analysis of the dependence of the positional angle of
dynamical axis and of the inclination on radius gives
the information about features of the gas motion in the galaxy.

The long-slit observations of NGC~7465 were performed
in two spectral ranges: ``green'' and ``red''.
In the ``green'' range containing the $\textrm{H}\beta$,
[$OIII$]$\lambda\lambda$4959,5007~\AA~ emission lines
and the absorption lines of the old stellar population
MgI 5175\AA, $\textrm{FeI}+\textrm{Ca~}5270$~\AA~
and others, we used the VPHG2300G grism with the
spectral resolution $\delta\lambda = 2.2$~\AA~ in
the range $\Delta\lambda$ = 4800{-}5570\AA.
The VPHG1200R grism was used for the ``red'' range
$\Delta\lambda$ = 5700-7400\AA~(the vicinity of the
H$\alpha$ line), and the spectral resolution was equal
to 5~\AA. The processing of the obtained data was
performed using standard procedures of the ESO-MIDAS
package. After the primary reduction, the smoothing
along the slit with a rectangular window 3 pixel in
height was carried out to increase
the signal-to-noise ratio. Line-of-sight velocities
of the gaseous component were measured from positions
of centers of Gaussians fitted to emission lines.
An accuracy of these measurements was estimated by
the night-sky HgI~$\lambda$5461~\AA ~and
[OI]~$\lambda$6300~\AA ~lines and was 10-15 km s$^{-1}$.
The cross-correlation method (Tonry, Davis, 1979)
was used to determine line-of-sight velocities and
velocity dispersions from absorption lines.

MPFS simultaneously takes spectra from 256 spatial elements
(in the form of square lenses) that form a
$16 \times 16$-element array in the plane of the sky.
During our observations, the angular size of one element
was $1''$. The spectral resolution was 3~\AA. The comparison
spectrum of the He-Ne-Ar lamp was used for the wavelength
scale calibration; the linearization accuracy was
${\sim}0.3$~\AA. The processing of the MPFS spectra was
carried out using a software package developed by
V.L.~Afanasiev and A.V.~Moiseev (SAO RAS).

\section{Multicolor photometry}

\begin{figure}[h!]
    \vspace*{-0.0cm}
    \hspace*{-0.0cm}
    \vbox{ \includegraphics{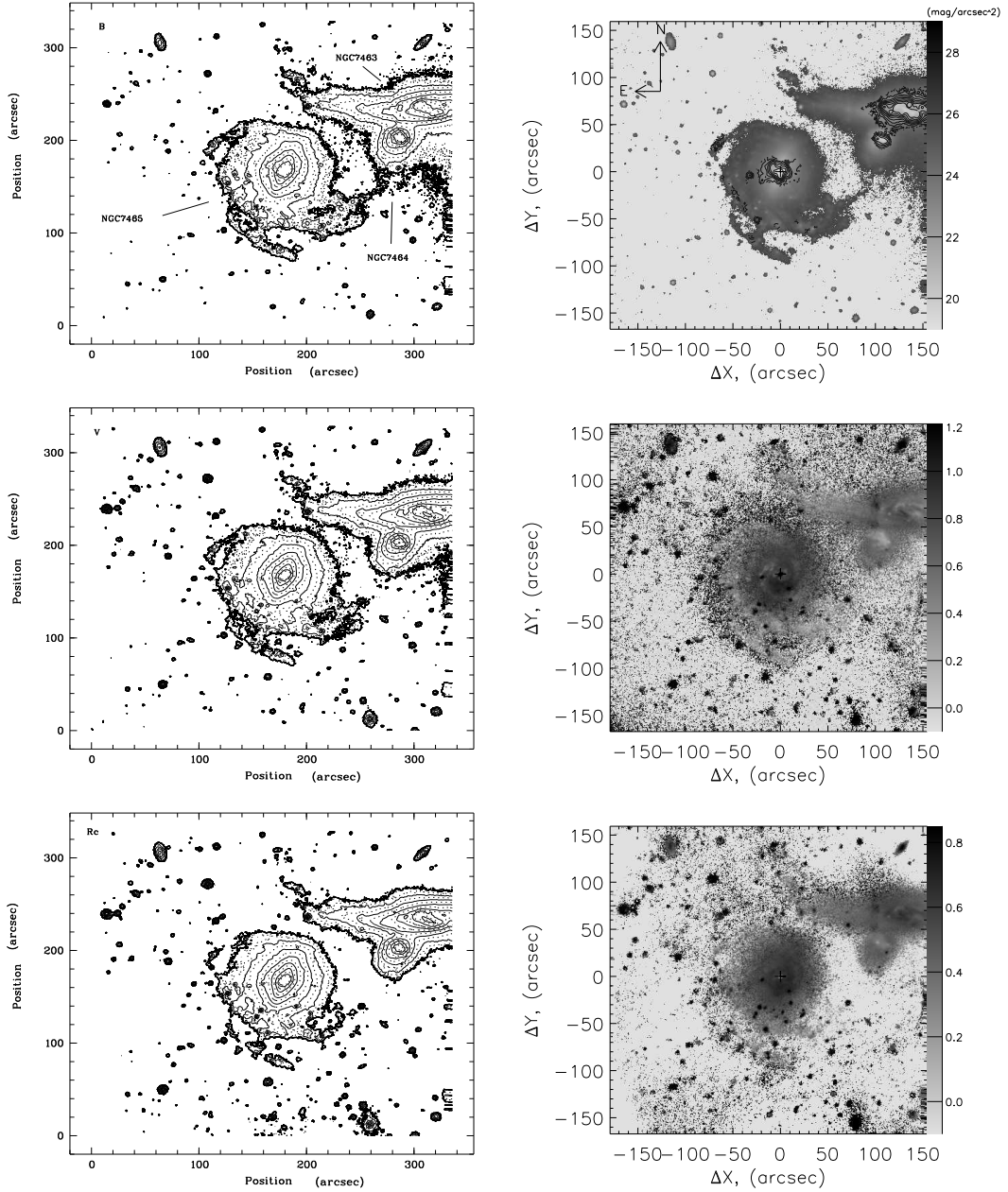}
             } \par
\vspace*{16.cm} \hspace*{-0.0cm}
\caption{Left column top-down:
isophotes of the galaxies in the $B$, $V$, $R_{\textrm{c}}$
bands (with a step equal to 0.5 mag/$\Box''$; the most outer
isophote in the $B$ band corresponds to the surface
brightness of 26~mag, that in $V$ corresponds to 25.5~mag,
and that in $R_{\textrm{c}}$ corresponds to 25~mag).
Right column top-down: the image in the $B$ band with
the superimposed isophotes in the H$\alpha$ line,
$B{-}V$ and $V{-}R_{\textrm{c}}$ color distribution maps.}
\label{f:7465BVR}
\end{figure}

During photometric observations, the task was to reveal
faint tidal structures between the galaxies of the triplet
and to study the expected polar ring, so a frame was
centered on NGC~7465, and the most distant galaxy
NGC~7463 is not whole in the frame. The image of
NGC~7465/64/63 in the $B$ band with the superimposed
isophotes in the H$\alpha$ line and the isophotes in
the $B$, $V$, $R_c$ bands are presented in
Fig.~\ref{f:7465BVR}. We obtained deeper images than in earlier works
(van Driel et al., 1992; Schmitt, Kinney, 2000), and in
our images the features of outer regions of galaxies
are seen more clearly.

Despite the ragged structure of the NGC~7465 galaxy,
three spiral arms and not a faint ring, the presence of
which was pointed out earlier by several authors (see,
e.g., van Driel et al., 1992, etc.), are prominent.
The northeastern and southeastern arms are superimposed
on the arc-like structure observed in HI and called
the polar ring by Li and Seaquist (1994). The southwestern
arm bends and extends to the northwest in the direction
of the NGC~7464 galaxy, whose isophotes are asymmetric
and stretched in the direction of this arm. Three spirals
manifest themselves in the color index distribution
too (Fig.~\ref{f:7465BVR}). However, only the brightest parts of the
spirals that are close to the main body of the galaxy
are seen because of the ragged structure and the low
surface brightness at the ends of spirals.

As noted above, NGC~7463 is not whole in the frame,
but in its images a central bar-like structure and
an eastern spiral arm, whose end turns to the north,
are clearly seen.

The total apparent magnitudes in three bands and
the integrated color indices of NGC~7465 and NGC~7464
were obtained using the multiaperture photometry method,
whose accuracy (as it was mentioned above) was
${\pm}0\overset{m}{.}1$. But emission lines find
themselves within the bandwidth of all three filters,
and this can lead to the increase of errors,
especially in regions where the contribution of
the radiation in emission lines is considerable.
The total magnitudes corrected for the extinction
in our Galaxy (Schlegel et al., 1998)
are given in Table~\ref{t:data7465}. Color indices of NGC~7464
are close to values typical for IrrI galaxies,
and color indices of NGC~7465 are close to
values typical for Sa-Sab galaxies.

The $B{-}V$ и $V{-}R_{\textrm{c}}$ color distributions
are presented in Fig.~\ref{f:7465BVR}. As follows from these maps,
the bluest color indices are observed in the circumnuclear
region of NGC~7464 ($B{-}V \approx 0\overset{m}{.}2$)
and in the region of spiral arms of NGC~7465
(${\approx}0\overset{m}{.}36$) and NGC~7464
(${\approx}0\overset{m}{.}3$). In the circumnuclear
region of NGC~7465, values of color indices are
distorted due to presence of emission lines.
At a distance of from $10''$ to $30''$ from the nucleus,
$B{-}V$ changes little (0$^m$.7--0$^m$.8);
then it gradually turns blue and reaches 0$^m$.36
in the region of spiral arms. The color index distribution
in the main body of NGC~7463 is extremely nonuniform which
seems to be connected with the presence of the bar,
numerous HII-regions, and the dust lane, where $B{-}V$
is $\approx$0$^m$.8.

\section{Morphology and structure of the triplet galaxies}

For the analysis of the photometric structure of the galaxies,
we used a technique proposed by Jedrzejewsky (1987).
It is based on the Fourier expansion of the deviation of
isophotes from the elliptic shape. For this purpose,
the IRAF system was used for the data processing.
Let us consider results for each galaxy that were obtained by us.

\begin{figure}[h!]
    \vspace*{-0.0cm}
    \hspace*{-0.0cm}
    \vbox{\includegraphics{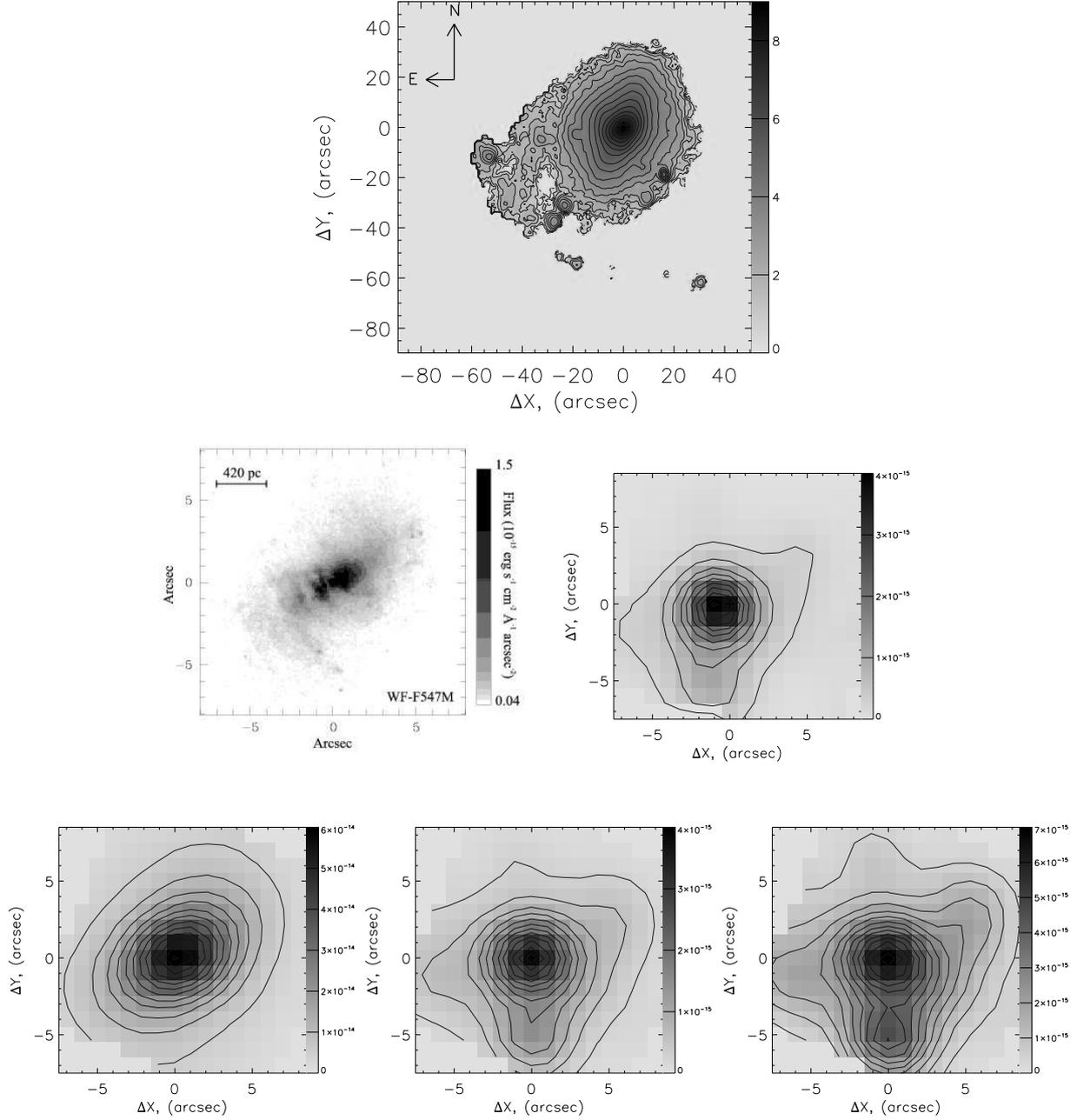}
             } \par
\vspace*{16.5cm} \hspace*{-0.0cm}
\caption{NGC~7465. Top: the image in the continuum
near H$\alpha$ according to the IFP data. Middle:
left: the image in the continuum in the F547 filter
(Fig.~24 (top, right) from Ferruit et al. (2000));
right: the image in the [OIII]~$\lambda$5007~\AA~
line according to the MPFS data. Bottom from left
to right: images in the continuum near H$\alpha$,
in the [NII]~$\lambda$6584~\AA~ line, and in the
H$\alpha$ line according to the MPFS data.}
\label{f:7465mpfs}
\end{figure}

\textbf{NGC~7465.} In the central region ($r \leq 5''$),
isophotes in the $R_{\textrm{c}}$ band have approximately
elliptic shape with the ellipticity $\epsilon \approx 0.2$
and the position angle of the major axis $\textrm{PA}
\approx 120^{\circ}$. We also analyzed the shape of
isophotes in this region in ``red'' and ``green'' continua
from the IFP and MPFS data (Fig.~\ref{f:7465mpfs}), in order to eliminate
the influence of emission lines. It turned out that values
of both $\epsilon$ and PA coincide with corresponding values
of the broadband photometry.

Images in the circumnuclear region of the galaxy in continua
(F547M and F791M filters) were obtained with the wide-field
camera (WFPC2) of the Hubble Space Telescope (HST)
(Ferruit et al., 2000). Thanks to the high spatial
resolution (of the order of $0.1''$), a series of
interesting details manifest itself in these images.
Figure~\ref{f:7465mpfs} presents a HST frame in the F547M filter.
Regions brightest in the continuum are located within
a circle with $r = 4''$ and display an inverted $S$-shaped
morphology with the position angle of the major axis
$\textrm{PA} \approx 120^{\circ}$; then the major axis
turns gradually, and PA $\approx 160^{\circ}{-}165^{\circ}$
at a distance of $10''$. The morphology of central regions
to the southeast of the nucleus is distorted strongly
by dust lanes perpendicular to the major axis of
the galaxy. A deep compact minimum in
the log $\textrm{(F547M/F791M)} ratio (\simeq{-}0.15$)
suggestive of the strong reddening corresponds to the nucleus.

According to our data (both the continuum and
broadband observations), $\epsilon$ increases with
the increase of the distance from the center of the
galaxy from 0.2 ($r = 2''$) to ${\sim}0.35$ (at $r =
10''{-}30''$) and decreases gradually to 0.2 in the
region of spiral arms; PA increases from $120^{\circ}$
to $160^{\circ}$ at $r$ from $2''$ to ${\sim}15''$ and
is constant further. Such behavior of PA agrees with
the HST data and with earlier ground-based observations
in the continuum (van Driel et al., 1992; Mulchaey et al.,
1996). The change of ellipticity in outer ($r > 30''$)
parts of the galaxy is probably connected with
the violation of the symmetry of spiral arms relative
to the center owing to the interaction.

We also carried out the analysis of the photometric
structure of NGC~7465 in the $K_s$ band, using data
of the Two Micron All Sky Survey (2MASS). The galaxy
disk radius in the image in this band is ${\approx}23''$.
In the central region $\textrm{PA} \approx 120^{\circ}$;
with the distance from the center, PA increases up
to $160^{\circ}$ (at $r = 15''$) and then it remains
approximately constant. The ellipticity changes from
0.2 to ${\approx}0.4$, and this coincides with
the behavior of these parameters in the optical range.

The brightness distribution of NGC~7465 in the
$R_{\textrm{c}}$ band was used for the decomposition
into components. Two regions can be singled out in
the profile along $\textrm{PA} = 160^{\circ}$:
the central one (up to $r \approx 5''{-}7''$ from
the center) with a steeper shape of the profile and
an extended outer one ($7''\leq r \leq 40''$) where
the profile is less steep. The brightness profile in
the outer region is well represented by the exponential
law with a scale factor
$h_{\textrm{d}} = 7\overset{''}{.}4$ and a cental
surface brightness $\mu_{\textrm{d}} = 18\overset{m}{.}0$,
along the almost whole length. After the subtraction
of an exponential disk with the values of $h_{\textrm{d}}$
and $\mu_{\textrm{d}}$ that were found above,
at $\textrm{PA} = 160^{\circ}$ and $i = 50^{\circ}$
taken from the observed brightness distribution,
the central structure stretched in the direction of
$120^{\circ}$ is seen more clearly. We tried to
approximate the photometric cut along this direction
by the Sersic law. The best agreement with the
observed profile is reached at the following parameters:
the index $n = 1 \pm 0.2$ (such index is typical
for disks), the effective radius $R_{e,b} = 4\overset{''}{.}6$,
and the central surface brightness $\mu_b = 16\overset{m}{.}3$.

Let us proceed to the analysis of the brightness
distribution in emission lines. The comparison of
brightness distributions in the continuum and in
H$\alpha$ (Fig.~\ref{f:7465mpfs} and \ref{f:7465BVR})
showed that they differ
noticeably. Isophotes in H$\alpha$ are stretched
along the NE-SW direction ($\textrm{PA} \sim50^{\circ}$),
and their shape is far from elliptic one (according
to a rough estimation, $\epsilon$
${\sim}0.3$ ($i \sim46^{\circ}$)). The radiation
maximum corresponds to the galaxy nucleus, and
a slightly less bright region is south of the
nucleus at $r \approx 5''$; other less bright
condensations are also observed. On the SE side,
an arched outgrowth with the brightening at its
end which apparently relates to the SE spiral arm
is well noticeable. A string of regions lighting
in the H$\alpha$ line is singled out on the southeast
side at a distance of $55''{-}70''$ from the galaxy
center. This string looks like a ``ragged'' semiring.
One more, extended region bright in H$\alpha$ is seen
to NE at a distance of ${\sim}80''$ from the center.
Let us note that all these regions fall on spiral arms
of the galaxy. A total flux in the в H$\alpha$ line
and a lower estimate of the star formation rate
(with no account taken of the extinction) that was
obtained on the basis of the relation from Kennicutt
(1998) are given in Table~\ref{t:data7465}.

The comparison of brightness distributions in the
permitted (H$\alpha$,
$\textrm{H}\beta$) and forbidden ([NII], [SII], [OIII])
emission lines (Fig.~\ref{f:7465mpfs}) for the central region of the
galaxy shows that images in all emission lines are similar
and there are no considerable differences in the extent
of the emission.

\begin{figure}[h!]
    \vspace*{-0.0cm}
    \hspace*{-0.0cm}
    \vbox{ \includegraphics{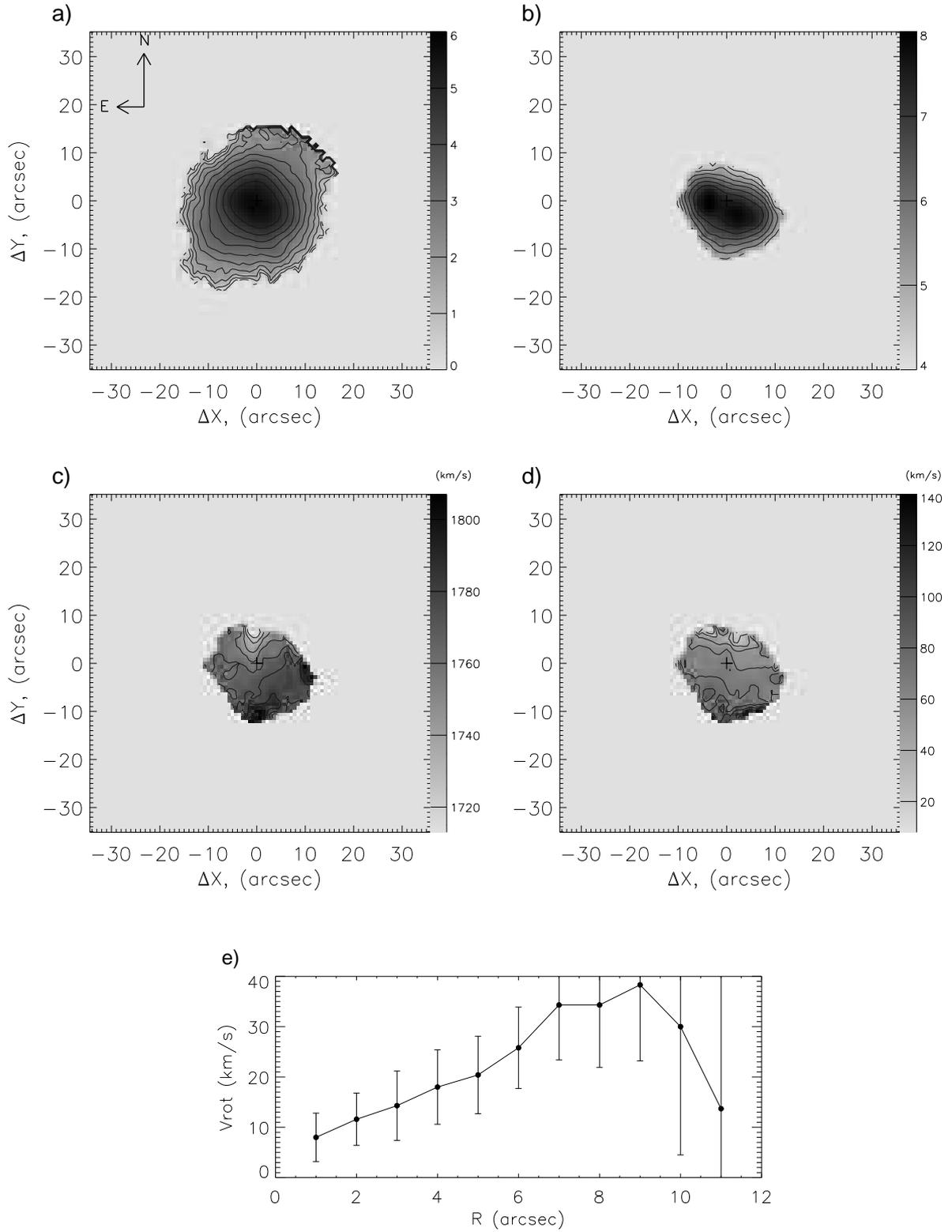}
             } \par
\vspace*{19.5cm} \hspace*{-0.0cm}
\caption{NGC~7464. Results from the IFP data:
(a) --- the brightness distribution in the narrow
continuum near the H$\alpha$ line; (b) ---
the brightness distribution in the H$\alpha$ line;
(c) --- the line-of-sight velocity field in the
H$\alpha$ line; (d) --- the line-of-sight velocity
dispersion field in the H$\alpha$ line. Results of
the ``tilted-ring'' method: (e) --- the rotation
curve of the gas.}
\label{f:n7464_ifp_ha}
\end{figure}

\textbf{NGC~7464.} The brightness distribution both in
broadband filters and in the continuum
(Fig.~\ref{f:7465BVR}, \ref{f:n7464_ifp_ha}) is
amorphous, and the shape of isophotes is approximately
elliptic. In the central region ($r \leq 7''$),
$\textrm{PA} \sim50^{\circ}$ and the ellipticity changes
from 0.3 ($r \approx 2''$) to 0.1 ($r \approx 8''$).
At a distance of ${\approx}9''{-}11''$, isophotes are round,
then $\epsilon$ increases up to 0.2 ($r \approx 15''$);
PA in outer parts is ${\sim}120^{\circ}$. The photometric
profile along the major axis ($\textrm{PA} = 120^{\circ}$)
is asymmetric; the SE-wing is flatter. The stretching
in the direction of the SW spiral arm of NGC~7465 increases
with the distance from the center, as we have already
mentioned in the section ``Multicolor Photometry''.
The turn of PA from ${\approx}50^{\circ}$ in the center
to ${\approx}120^{\circ}$ in outer parts is observed.

Brightness distributions in the H$\alpha$ and
[NII]~$\lambda$6584~\AA\ lines are similar, therefore
in Fig.~\ref{f:n7464_ifp_ha} we give only the image in the H$\alpha$.
In the central part, two bright regions located
approximately along the direction of $65^{\circ}$
on both sides from the nucleus and embedded in to
the diffuse radiation in these lines are prominent.
It was noted in van Driel et al. (1992) that there
is possibly a bar in the center of this galaxy;
but our study of the kinematics (see the following
section) does not confirm the presence of the bar.
PA of outer isophotes ($r \geq 8''$) is
${\approx}68^{\circ}$. We should note that isophotes
in emission lines are turned relative to isophotes
in the continuum.

\begin{figure}[h!p]
    \vspace*{-0.0cm}
    \hspace*{-0.0cm}
    \vbox{ \includegraphics{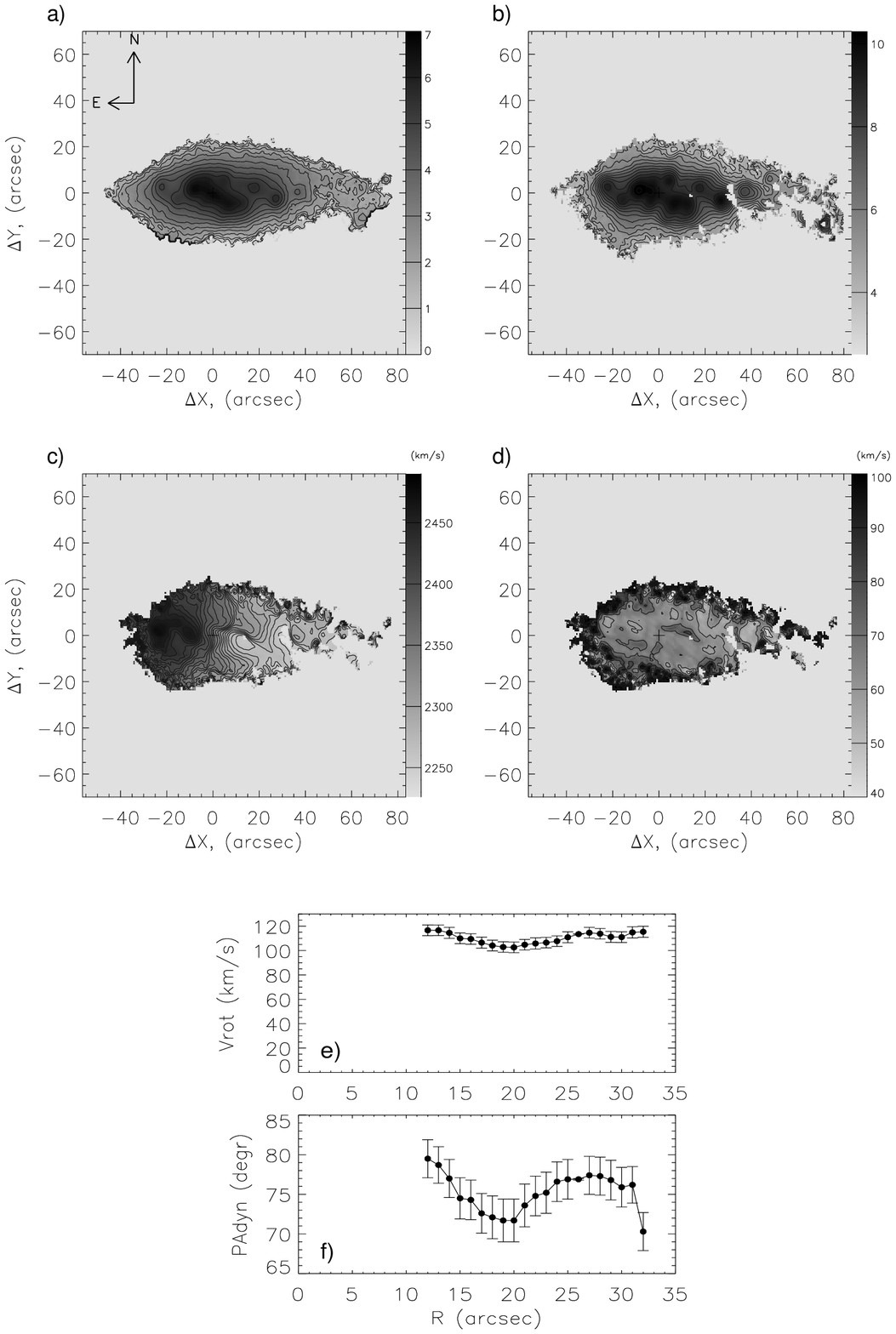}
             } \par
\vspace*{21.cm} \hspace*{-0.0cm}
\caption{NGC~7463. Results from the IFP data: (a) ---
the brightness distribution in the narrow continuum
near the H$\alpha$ line; (b) --- the brightness
distribution in the H$\alpha$ line; (c) --- the
line-of-sight velocity field in the H$\alpha$ line;
(d) --- the line-of-sight velocity dispersion field
in the H$\alpha$ line. Results of the ``tilted-ring''
method: (e) --- the rotation curve of the gas;
(f) --- PA$_{\textrm{dyn}}(R)$. }
\label{f:n7463_ifp}
\end{figure}

\textbf{NGC~7463.} Figures~\ref{f:7465BVR} and \ref{f:n7463_ifp}a
present brightness
distributions in broadband filters and the continuum,
respectively. Main parameters of the bar are:
$\textrm{PA}_{\textrm{bar}} \approx 57^{\circ}$,
the ellipticity of bar is $\epsilon_{\textrm{bar}}
\approx 0.7$, and the size of the semimajor axis of
bar $a_{\textrm{bar}}$ is ${\approx}11''$. The disk
of the galaxy has the following characteristics:
PA$_{\textrm{disk}} \approx 90^{\circ}$, and
$\epsilon_{\textrm{disk}}$ changes smoothly from
0.45 at a distance of 12$''$ to 0.6 at $r = 30''$
(the inclination of the disk changes from $57^{\circ}$
to $66^{\circ}$). Outer parts of the disk are
asymmetric relative to the minor axis; the western
part looks more disturbed, and the turn/bend of
this part of the disk to the south is observed.

In the image of the galaxy in the H$\alpha$ line
(Fig.~\ref{f:n7463_ifp}b), the bar-like structure is not singled out,
but only two brightenings at its ends are seen, and
the brightness of the eastern one is higher.
In addition, several bright condensations located
along spiral arms are observed in the given image,
and they are probably HII-regions. All this is
embedded into the diffuse radiation in the H$\alpha$
line that extends approximately to the same distances
as the radiation in the continuum.

\section{Kinematics of gas and stars}

\begin{figure}[h!p]
    \vspace*{-0.0cm}
    \hspace*{-0.0cm}
    \vbox{ \includegraphics{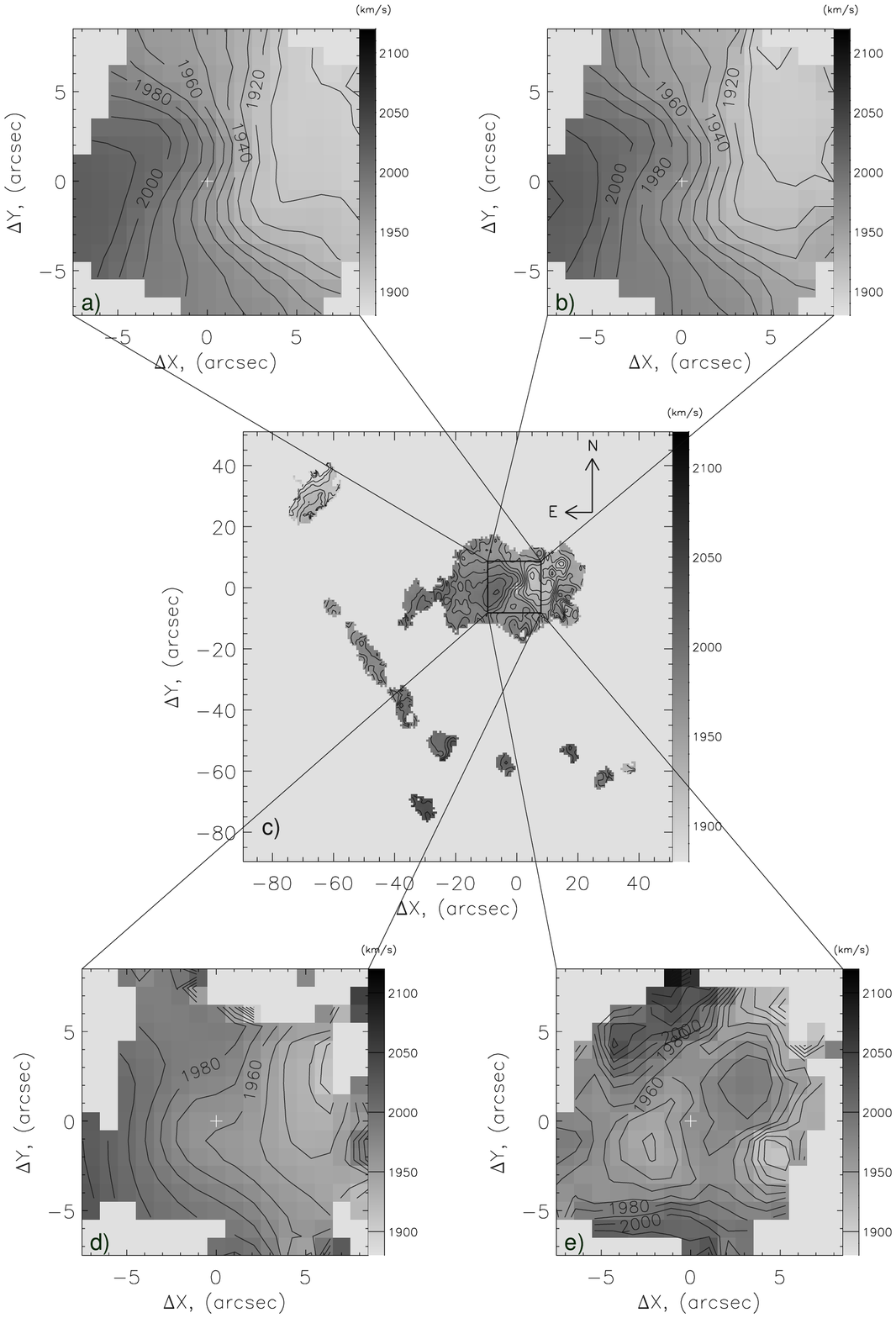}
             } \par
\vspace*{21.5cm} \hspace*{-0.0cm}
\caption{NGC~7465. Line-of-sight
velocity fields of the ionized gas in the following
lines: (a) --- H$\alpha$ (MPFS); (b) ---
[NII]~$\lambda$6584~\AA~ (MPFS); (c) ---
H$\alpha$ (IFP); (d) --- [OIII]~$\lambda$5007~\AA~
(MPFS). (e) --- the line-of-sight velocity field
of the stellar component (MPFS). }
\label{f:n7465_ifp}
\end{figure}

Spectral observations of the triplet galaxies included
observations with IFP. On their basis, line-of-sight
velocity fields and velocity dispersion maps in the
H$\alpha$ line for each of galaxies (Fig.~\ref{f:n7465_ifp},
\ref{f:n7464_ifp_ha}c,d,
\ref{f:n7463_ifp}c,d)
as well as in the [NII]~$\lambda$6584~\AA~ line for
NGC~7464 were constructed. For NGC~7465, the more detailed
spectral study was performed (see Table~\ref{t:zhurnal_spectr}).
Observations
with the MPFS and long-slit spectrograph were carried
out in two spectral regions (``red'' and ``green''),
and this allowed us to study the kinematics not only
of the gaseous component but also of the stellar component.

\textbf{NGC~7465.} Let us consider the results of
observations of the gaseous component of this galaxy.
Figure~\ref{f:n7465_ifp}c presents a large-scale line-of-sight velocity
field in the H$\alpha$ line. It is obvious from the
figure that the gas disk rotates, and the eastern side
approaches us, while the western side recedes from us.
In the central region of the galaxy ($r < 4''{-}5''$),
a knee of isovels is noticeable. We will dwell on this
feature below when discussing the MPFS data. At greater
distances (up to $r \approx 20''{-}25''$), the shape of
isovels corresponds to the rotation of a disk.

\begin{figure}[h!p]
    \vspace*{-0.0cm}
    \hspace*{-0.0cm}
    \vbox{\includegraphics{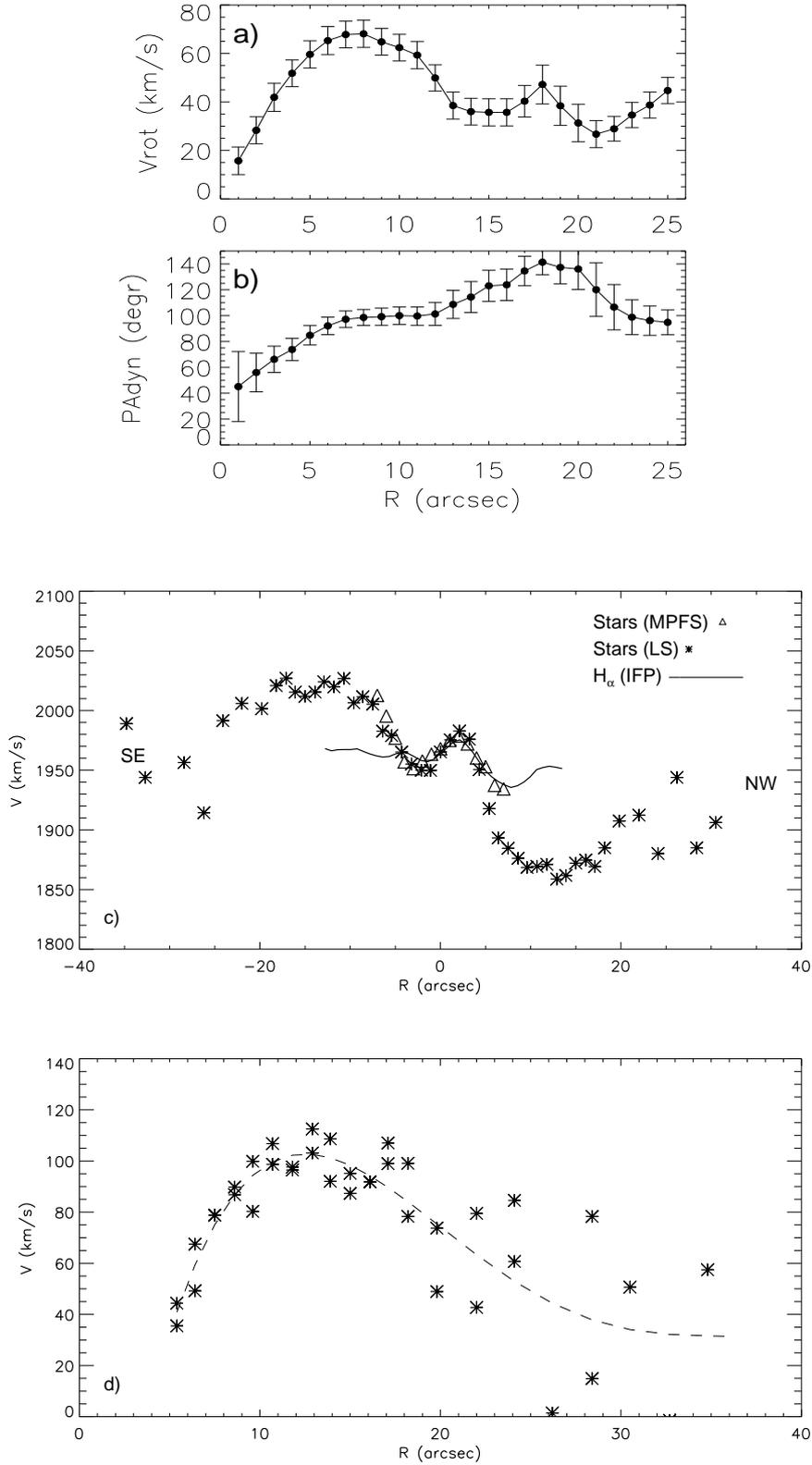}
             } \par
\vspace*{21.cm} \hspace*{-0.0cm}
\caption{NGC~7465. Results of the ``tilted-ring'' method:
(a) --- the rotation curve of the ionized gas, (b) ---
PA$_{\textrm{dyn}}(R)$. (c) --- line-of-sight velocity
curves of stars (slit and MPFS) and of the ionized gas
(IFP) along $\textrm{PA} = 160^{\circ}$; (d) --- the
rotation curve of stars ($\ast$ --- the observed values
of the rotation velocity, dashed line --- the average
smooth rotation curve).}
\label{f:n7465_vrot}
\end{figure}

The analysis of the large-scale line-of-sight velocity
field by the ``tilted-ring'' method has shown that the
photometric and dynamic centers coincide and the
heliocentric velocity of the system is 1963~km/s.
Figure~\ref{f:n7465_vrot}a shows a rotation curve of the ionized gas;
the maximum rotation velocity is reached at the
distance of 8$''$ and is equal to 70~km/s, then it
decreases, and in the region from 13$''$ to 25$''$
it changes little and is equal approximately to 40~km/s.
The change of the position angle of the dynamic axis
with the distance from the center is given in
Fig.~\ref{f:n7465_vrot}b.
It is obvious from this figure that the dynamic axis
turns smoothly with distance from the center, and
the inclination of the disk to the plane of the sky
changes too. Parameters of the model of circular
rotation of the ionized gas turn out to be
the following: at $r = 2''{-}3''$
$\textrm{PA}_{\textrm{dyn}}\approx 50^{\circ}$ and
the inclination to the plane of the sky is
$i_{\textrm{dyn}}\approx 50^{\circ}$; while in
the region from 15$''$ to 20$''$ PA$_{\textrm{dyn}}$
changes from $120^{\circ}$ to $130^{\circ}$ and
$i_{\textrm{dyn}} \approx 60^{\circ}$.

For the central region of NGC~7465, velocity fields
in all emission lines were constructed on the basis
of the MPFS data. They turned out to be similar,
therefore fields in the H$\alpha$, [NII] and [OIII]
lines are given in Fig.~\ref{f:n7465_ifp}a,b,d.
In the circumnuclear
region ($r \leq  3 ''$), a knee of isovels is
observed, like according to the IFP data.
As a whole the eastern part of the gas disk
recedes from us, while the western part approaches us;
the dynamic axis turns gradually with the increase
of distance from the center from
$\textrm{PA}_{\textrm{dyn}}\approx 70^{\circ}$
($r = 2''$) to $\textrm{PA}_{\textrm{dyn}}
\approx 110^{\circ}$ ($r = 8''$), and this coincides
with data obtained with the IFP, within the errors limits.

As concerns bright condensations in H$\alpha$ that
form the arc/semiring (see~Fig.~\ref{f:n7465_ifp}c), then
(as we have mentioned when analyzing structures
of this galaxy) the majority of them belong to
different spiral arms (NE-, SE-, SW-arms).
Their velocities most likely characterize line-of-sight
velocities of gas in corresponding parts of spirals.

Line-of-sight velocity curves in emission lines
(long-slit spectra) along major axes of the main body
($\textrm{PA} = 160^{\circ}$) and of the supposed
polar ring ($\textrm{PA} = 45^{\circ}$) are similar
to each other and agree well with the MPFS and IFP data.
The coincidence of line-of-sight velocity values
measured using three different spectral instruments,
within the accuracy, indicates the reliability of our data.

Let us proceed to the consideration of kinematics of
the stellar component. Figure~\ref{f:n7465_ifp}e presents the velocity
field of stars that was obtained from observations
in the ``green'' range with MPFS. In the central region
($r \leq 4''{-}5''$), the shape of isovels is regular
and typical for a rotating disk. The analysis of this
field by the ``tilted-ring'' method has shown that
$\textrm{PA}_{\textrm{dyn},\textrm{st}} \approx
300^{\circ}$ and $i_{\textrm{dyn},\textrm{st}}
\approx 60^{\circ}$ and they are close to photometric
parameters obtained by us in the previous section.

Let us consider the motion behavior of the stellar
component of this galaxy according to the long-slit
spectroscopy data. Figure~\ref{f:n7465_vrot}c presents
the line-of-sight
velocity curve of stars along the major axis of
the galaxy ($\textrm{PA}
= 160^{\circ}$). Absorption lines are seen up to
30$''$ from the center in this direction. Values of
line-of-sight velocities of stars that were derived
from observations with the long-slit spectrograph and
MPFS coincide within the error limits. The change of
the gradient direction to the opposite one is observed
in the line-of-sight velocity curve of stars at $r \geq
6''$ (outside the MPFS's field of view) (Fig.~\ref{f:n7465_vrot}c).
This is probably connected with the rotation of the
main stellar disk of the galaxy around its minor axis;
the NW-side of the disk approaches us, while
the SE-side recedes from us.

The comparison of the line-of-sight velocity curve
of stars at $\textrm{PA} =
160^{\circ}$ with a cut of the line-of-sight velocity
field in the H$\alpha$ line along the same direction
(Fig.~\ref{f:n7465_vrot}c) shows that line-of-sight velocities of
stars and of the ionized gas differ considerably
starting from the distance of ${\approx}3''{-}4''$
from the center. This means that the rotation of
these components in the given region occurs around
different axes.

If we assume that the dynamic axis of the stellar
disk in outer regions of the galaxy coincides with
the photometric axis, then
$\textrm{PA}_{\textrm{dyn},\textrm{st}} \approx
160^{\circ}$ and the inclination of the stellar
disk to the plane of the sky is
$i_{\textrm{dyn},\textrm{st}} \approx 50^{\circ}$.
Under these assumptions we can construct the rotation
curve of the stellar disk of the galaxy (Fig.~\ref{f:n7465_vrot}d).
The maximum rotation velocity is reached at the
distance $r_{\textrm{max}} = 13''$ from the nucleus,
and it is equal to ${\approx}105$~km/s.

On the basis of the study of the kinematics of
stars we ascertained that in the central region of
NGC~7465, a stellar disk-like structure is singled out,
and its rotation axis is at a considerable angle
(${\sim}140^{\circ}$) to the rotation axis of the
main stellar disk.

Let us say a few words about line-of-sight velocity
dispersions of the stellar and gas components.
According to our data, the velocity dispersion of
both stars and the ionized gas is small and does not
considerably exceed determination errors. A small
increase of the dispersion by 10--20~km/s is observed
in the circumnuclear region ($r \leq 2''$) for both components.

\textbf{NGC~7464.} For this galaxy, line-of-sight
velocity fields in the H$\alpha$ and
[NII]$\lambda$6584\AA~ lines were constructed;
they turned out to be similar, therefore the field
in the H$\alpha$ line is given in Fig.~\ref{f:n7464_ifp_ha}c.
Although the shape of isovels is not very smooth,
the rotation of the galaxy with a small velocity
is noticeable (the N-side recedes from us, and
the S-side approaches us). A nucleus is clearly
seen in the galaxy image in the continuum.
The line-of-sight velocity at this point is ${\approx}
1765$~km/s, and it is taken for the velocity of
the system. It coincides with results of
van Driel et al. (1992), but differs from neutral
hydrogen data (Paturel et al., 2003; see Table~\ref{t:data7465})
by 100~km/s. The asymmetry between N- and S-parts
of the velocity field is observed: the S-part is
more extended and is stretched in the direction of
the spiral arm of the NGC~7465 galaxy. In addition,
PA$_{\textrm{dyn}} = 240^{\circ}$ in the region
$r \leq 5''$, while $\textrm{PA}_{\textrm{dyn}} =
185^{\circ}$ in more outer parts. Under the assumption
of the circular rotation, we constructed the rotation
curve (Fig.~\ref{f:n7464_ifp_ha}e) at the following parameters:
$\textrm{PA}_{\textrm{dyn}} = 185^{\circ}$,
$i_{\textrm{dyn}} = 40^{\circ}$, and
$V_{\textrm{sys}} = 1765$~km/s. The maximum value
of the rotation velocity is reached at a distance
of 9$''$ and is equal to 40~km/s. Despite
determination errors of these values, one can
assert that NGC~7464 has the regular rotation.

The increase of the velocity dispersion from the
N-edge of the galaxy to the S-edge by approximately
50~km/s is noticeable in the velocity dispersion
map (Fig.~\ref{f:n7464_ifp_ha}d); this increase can be connected with
the interaction with NGC~7465. Additional observations
are necessary for the refinement of the velocity
dispersion behavior.

\textbf{NGC~7463.} The line-of-sight velocity field
of this galaxy is rather complex (Fig.~\ref{f:n7463_ifp}c). It is known
that in the bar region, noncircular motions should be
observed, therefore we tried to analyze a part of the
line-of-sight velocity field of NGC~7463 outside the
bar ($12''\leq r \leq 34''$) by the ``tilted-ring''
method. The model of the circular rotation of disk with
the following parameters gave the best agreement with
the observed field: positions of the photometric and
dynamic centers coincide, the heliocentric velocity of
the system is 2366~km/s, PA$_{\textrm{dyn}}$
${\sim}75^{\circ}$, and under the assumption of
the thin disk with $i_{\textrm{dyn}} = 62^{\circ}$.
Figure~\ref{f:n7463_ifp}e presents the corresponding rotation curve.
In the region of $12''\leq r \leq 34''$,
the rotation velocity changes little and is equal
to ${\approx} 115$~km/s.

\section{Discussion of results and conclusion}

Photometric and spectral data obtained by us for
the NGC~7465/64/63 triplet galaxies showed the
interesting complex structure and kinematics in
each of them. Let us note right away that we did
not find the outer classical polar ring in NGC~7465
that was suspected by Whitmore et al. (1990).
According to our data, the observed SE-arc (see
images in H$\alpha$ (Fig.~\ref{f:7465BVR} right column on top))
consists of separate condensations belonging to
different spiral arms. As it has turned out,
the distribution and motion behavior of
the ionized gas differ from corresponding data
for the neutral hydrogen (Li and Seaquist, 1994);
the ionized gas of this galaxy most likely forms
its own detached system.

Moreover, such facts as the disturbed structure
of NGC~7464, the stretching of its outer isophotes
in the continuum and emission lines in the direction
of NGC~7465, and the spiral arm extending from
NGC~7465 to NGC~7464 led us the conclusion about
the existence of the close connection between
these galaxies. As a whole, photometric and
spectral properties of NGC~7464 that were
discovered by us (such as blue color indices;
the presence of regular motions typical for disks)
allowed us to relate it to the IrrI type.

In its turn, the whole set of our data for NGC~7463
shows that it is the barred spiral galaxy of
the SBb-c type. However, the bend of its outer
parts indicates that in the past, the close encounter
of NGC~7463 with a galaxy went by occurred;
it possibly was one of galaxies of the triplet or
of the NGC~7448 group.

Let us discuss at greater length the results of
the analysis of data for NGC~7465 that were
obtained by us. In addition to the main stellar
disk with the position angle of the major axis
of $160^{\circ}$ and with the inclination to
the plane of the sky of ${\sim}50^{\circ}$,
the distinct inner (circumnuclear) stellar
structure of the radius of ${\sim}4''$ (0.56~kpc)
with $\textrm{PA}_{\textrm{phot}} = 120^{\circ}$
was revealed in this galaxy. Photometric properties
(the shape of isophotes in images in different
spectral ranges, the approximation of
the photometric cut along its major axis by
the Sersic law with the index $n = 1 \pm 0.2$)
and the stellar kinematics in the indicated
region (the appearance of the velocity field
of stars corresponds to the rotation of
the disk with PA$_{\textrm{dyn},\textrm{st}}
= 300^{\circ}$ and $i_{\textrm{dyn},\textrm{st}}$
${\sim} 60^{\circ}$) led us to the conclusion
about the presence of the inner stellar disk
almost ``counter-rotating'' relative to
the main stellar disk.

On the basis of the large-scale brightness
distribution in the H$\alpha$ line and of
the complex behavior of isovels of
the ionized gas in the region of radius of
25$''$ (PA$_{\textrm{dyn},\textrm{gas}}$
changes from ${\approx} 70^{\circ}$ at $r =
2''$ to $\textrm{PA}_{\textrm{dyn},\textrm{gas}}\approx
120^{\circ}$ at $r = 20''$; $i_{\textrm{dyn},\textrm{gas}}
\sim 50^{\circ}$), we propose the presence of
the warped gas disk; in the circumnuclear region
($r \leq 5''$), this gas disk is polar relative
to the main stellar disk, and the estimation of
an angle between planes of the main stellar disk
and the gas disk (when calculating by its outer
border) gives values of $45^{\circ}$ and $83^{\circ}$.
The observed circumnuclear dust lanes perpendicular
to the major axis of the galaxy also speak in favor
of the polarity of the gas disk (see, e.g.,
Sil'chenko and Afanasiev, 2004).

The origin of the observed structure of NGC~7465
that combines, at the minimum, three systems
distinct by their properties (the inner and
main stellar disks~$+$~the warped gas disk) is
owed to the gravitational interaction with
other galaxies. It seems unlikely to us that
so complex structure was formed as a result of
a single interaction. The circumnuclear stellar
disk, most probably, could be formed as a result
of the capture and disruption of a dwarf companion.
The warped gas disk could arise in consequence of
the accretion of the matter from a gas-rich galaxy
onto NGC~7465 (Bournaud and Combes, 2003).
Judging by our results, NGC~7464 could serve
as such galaxy-companion.

The authors are grateful to the Large Telescope
Program Committee for the allocation of the observational
time at the 6-m telescope, to V.L.~Afanasiev (SAO RAS)
for the allocation of the Multi Pupil Fiber Spectrograph
for our observations, A.V.~Moiseev and S.N.~Dodonov
(SAO RAS) for the assistance in the carrying out of
observations at the 6-m telescope, and separately to
A.V.~Moiseev for the allocation of data analysis codes
and for valuable remarks in preparing the text of
the paper. O.A.~Merkulova is also grateful for
the support within the event 1.4 of the Federal Target
Program ``Kadry'' (contract no.~14.740.12.0852).

{\large \bf References}
\begin{flushleft}

1. V.L. Afanasiev, S.N. Dodonov, V.L. Moiseev,
\textit{Stellar Dynamics: from Classic to Modern}. (Ed. Ossipkov
L.P. and Nikiforov I.I., Saint Petersburg: St. Petersburg
University, 2001), p.103

2. V.L. Afanasiev and A.V. Moiseev, Astron. Lett. $\textbf{31}$, 194 (2005)

3. K.G. Begeman, Astron. Astrophys. $\textbf{223}$, 47 (1989)

4. F. Bournaud and F. Combes, Astron. Astrophys. $\textbf{401}$, 817 (2003)

5. C. Casini and J. Heidmann, Astron. Astrophys. Suppl. Ser.
$\textbf{34}$, 91 (1978)

6. M.X. Fernandez et al., Astron. J. $\textbf{139}$, 2066 (2010)

7. P. Ferruit, A.S. Wilson, and J. Mulchaey,
Astrophys. J. Suppl. Ser. $\textbf{128}$, 139 (2000)

8. R.I. Jedrzejewsky, Mon. Not. R. Astr. Soc. $\textbf{226}$, 747 (1987)

9. R.C. Kennicutt, Ann. Rev. Astron. Astrophys. $\textbf{36}$, 189 (1998)

10. A.U. Landolt, Astron. J. $\textbf{88}$, 439 (1983)

11. J.G. Li and E.R. Seaquist, Astron. J. $\textbf{107}$, 1953 (1994)

12. J.S. Mulchaey, A.S. Wilson, and Z. Tsvetanov, Astron. J. Suppl. Ser.
$\textbf{102}$, 309 (1996)

13. A.V. Moiseev, Bull. SAO $\textbf{54}$, 74, (2002)

14. A.V. Moisseev and V.V. Mustsevoi, Astron. Lett. $\textbf{26}$, 565 (2000)

15. S.I. Neizvestnyi, Izv. Spets. Astrofiz. Observ. $\textbf{17}$, 26 (1983)

16. G. Paturel, G. Theureau, L. Bottinelli, et al., Astron. Astrophys.
$\textbf{412}$, 57 (2003)

17. L.V. Shalyapina, et al., Astronomy Letters $\textbf{33}$, 520 (2007)

18. D.J. Schlegel, D.P. Finkbeiner, M. Davis,
Astrophys. J. $\textbf{500}$, 525 (1998)

19. H.R. Schmitt and A.L. Kinney,
Astrophys. J. Suppl. Ser. $\textbf{128}$, 479 (2000)

20. O.K. Sil'chenko and V.L. Afanasiev, Astron. J. $\textbf{127}$, 2641 (2004)

21. C.M. Springob, M.P. Haynes, P. Martha, et al.,
Astrophys. J. Suppl. Ser. $\textbf{160}$, 149 (2005)

22. B. Takase, Astron. Soc. of Japan. Publications, $\textbf{32}$, 605 (1980)

23. J. Tonry and M. Davis, Astron. Astrophys. $\textbf{84}$, 1511 (1979)

24. W. van Driel et al., Astron. Astrophys. $\textbf{259}$, 71 (1992)

25. B.C. Whitmore, R.A. Lucas, D.B. McElroy, et al.,
Astron. J. $\textbf{100}$, 1489 (1990)

\end{flushleft}

\end{document}